\documentclass[twocolumn,aps,pra,superscriptaddress,floatfix,longbibliography]{revtex4-1}

\usepackage{graphicx}
\usepackage{float}
\usepackage{bm,amsmath,amssymb}
\usepackage{color}
\usepackage{braket}
\usepackage{hyperref}
\usepackage[dvipsnames]{xcolor}
\hypersetup{
  colorlinks = false,
  urlcolor = black,
}
\usepackage{verbatim}
\usepackage{lipsum}
\usepackage{soul}

\newcommand{\pmh}[1]{\textcolor{black}{#1}}

\newcommand{\redline}[1]{\textcolor{black}{#1}}

\newcommand{\sm}{{\sigma_-}}

\newcommand{\sz}{{\sigma_z}}
\newcommand{\sx}{{\sigma_x}}
\newcommand{\sy}{{\sigma_y}}
\newcommand{\tsm}{{\tilde{\sigma}_-}}
\newcommand{\tsp}{{\tilde{\sigma}_+}}
\newcommand{\tsz}{{\tilde{\sigma}_z}}
\newcommand{\tsx}{{\tilde{\sigma}_x}}
\newcommand{\tsy}{{\tilde{\sigma}_y}}

\newcommand{\bdk}{{b_k^\dagger}}

\begin{document}
\title{Bath engineering of a fluorescing artificial atom with a photonic crystal}
\author{P. M. Harrington}
\email{patrick.harrington@wustl.edu}
\affiliation{Department of Physics, Washington University, St.~Louis, Missouri 63130}
\author{M. Naghiloo}
\affiliation{Department of Physics, Washington University, St.~Louis, Missouri 63130}
\author{D. Tan}
\affiliation{Department of Physics, Washington University, St.~Louis, Missouri 63130}
\affiliation{Shenzhen Institute for Quantum Science and Engineering and Department of Physics, Southern University of Science and Technology, Shenzhen 518055, People's Republic of China}
\author{K. W. Murch}
\email{murch@wustl.edu}
\affiliation{Department of Physics, Washington University, St.~Louis, Missouri 63130}
\affiliation{Institute for Materials Science and Engineering, St.~Louis, Missouri 63130}
\date{\today}
\begin{abstract}
{We demonstrate how the dissipative interaction between a superconducting qubit and a microwave photonic crystal can be used for quantum bath engineering. The photonic crystal is created with a step-impedance transmission line which suppresses and enhances the quantum spectral density of states, influencing decay transitions of a transmon circuit. The qubit interacts with the transmission line indirectly via dispersive coupling to a cavity. We characterize the photonic crystal density of states from both the unitary and dissipative dynamics of the qubit. When the qubit is driven, 
\redline{it dissipates into} the frequency dependent density of states of the photonic crystal. Our result is the deterministic preparation of qubit superposition states as the steady-state of coherent driving and dissipation near by the photonic crystal band edge, which we characterize with quantum state tomography. Our results highlight how the multimode environment from the photonic crystal forms a resource for quantum control.}
\end{abstract}
\maketitle

In experimental quantum information processing, there exists a tradeoff between control of quantum states and dissipation. However, dissipation can in fact be a resource for quantum control. An early example of dissipation engineering is laser cooling of atoms, where drive in combination with atomic decay is used to initialize atomic states \cite{wine79}. Such techniques have been extended to cool mechanical objects through cavity dissipation \cite{aspe14} and for control of quantum circuits {\cite{wils07,kapi17}}. Dissipation engineering with quantum circuits has been demonstrated for a variety of applications including state reset  \cite{geer13,bout17,wong19} and the creation of entangled states \cite{aron14,aron16,reit16,shan13,lin13,kimc16}. More generally, these are examples of quantum bath engineering, where decay is deliberately used as a means for quantum state preparation \cite{poya96,carv01,vers09,murc12,lu17,legh15,haco15,liu17,kapi17}. In another research arena, there is growing interest in the interaction of quantum systems with multimode environments \cite{sund15,mart18,kuzm18} such as photonic crystals \cite{cerb14,liu17,sund18,mirh18} where impurity models \cite{hur12,gold13}, ultrastrong coupling \cite{forn17, yosh17, mart18}, and driven-dissipative phase transitions \cite{houc12,raft14,fitz17} can play a leading role. In this letter, we show how the dissipation of a  quantum circuit into the multiple electromagnetic modes that form the  bands and gaps of a photonic crystal can be used to prepare non-trivial quantum states of the circuit through bath engineering. Our bath engineering protocol results in deterministic decay to superposition states of the two lowest energy levels of the superconducting circuit, which we investigate with full quantum state tomography. We find close agreement between experimentally measured steady-states and the predicted Lindblad evolution for a range of coherent drive parameters. Our results add to the growing tool box of dissipation engineering techniques for quantum control of superconducting circuits. 

\begin{figure}[h]
\centering
\includegraphics[width=8.6cm]{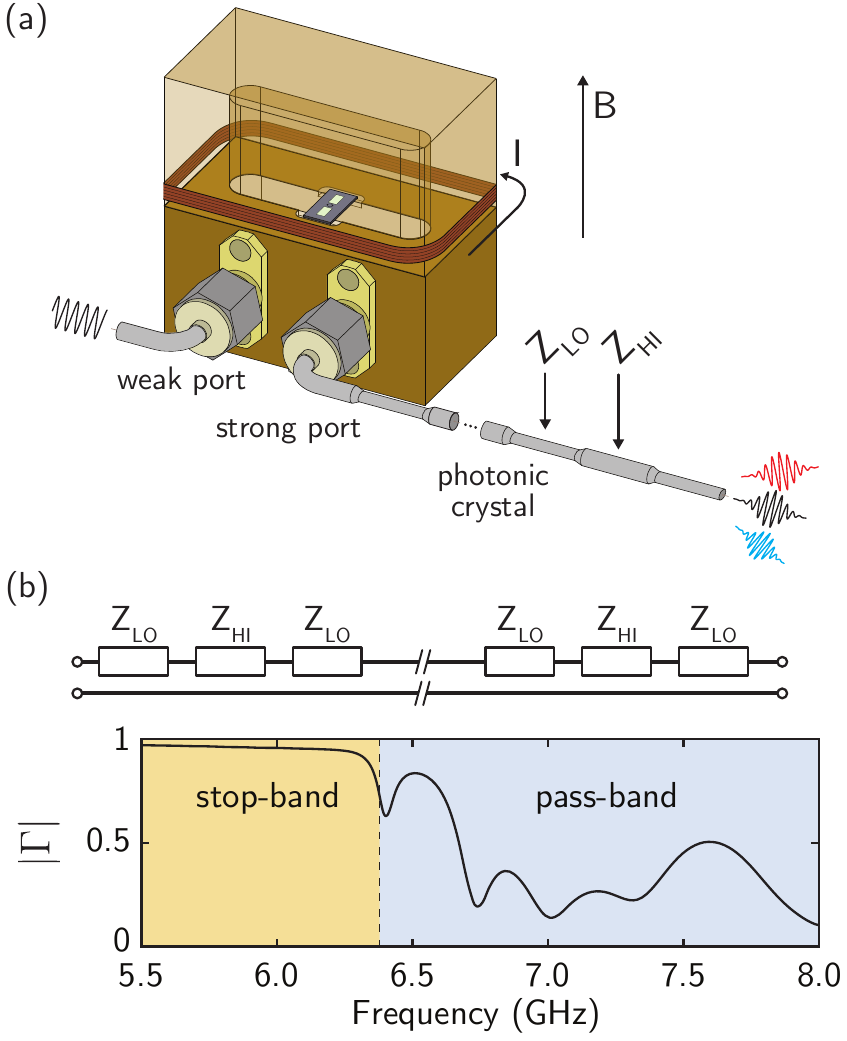}
\caption{(a) A transmon circuit is dispersively coupled to a copper waveguide cavity. The transmon junction has a superconducting quantum interference device (SQUID) geometry which allows for tuning of its resonant frequency. The cavity has a 
{weakly coupled antenna port for applying drive pulses to the qubit} and a second antenna port that is strongly coupled to a coaxial transmission line photonic crystal. (b) A standard $50\,\Omega$ coaxial cable is modified to have spatially periodic capacitive loading, which results in the opening of a photonic bandgap. The reflection coefficient magnitude $|\Gamma|$ of the photonic crystal measured at room temperature displays a frequency stop-band.}
\label{fig:fig1}
\end{figure}

\begin{figure}
\centering
\includegraphics[width=8.6cm]{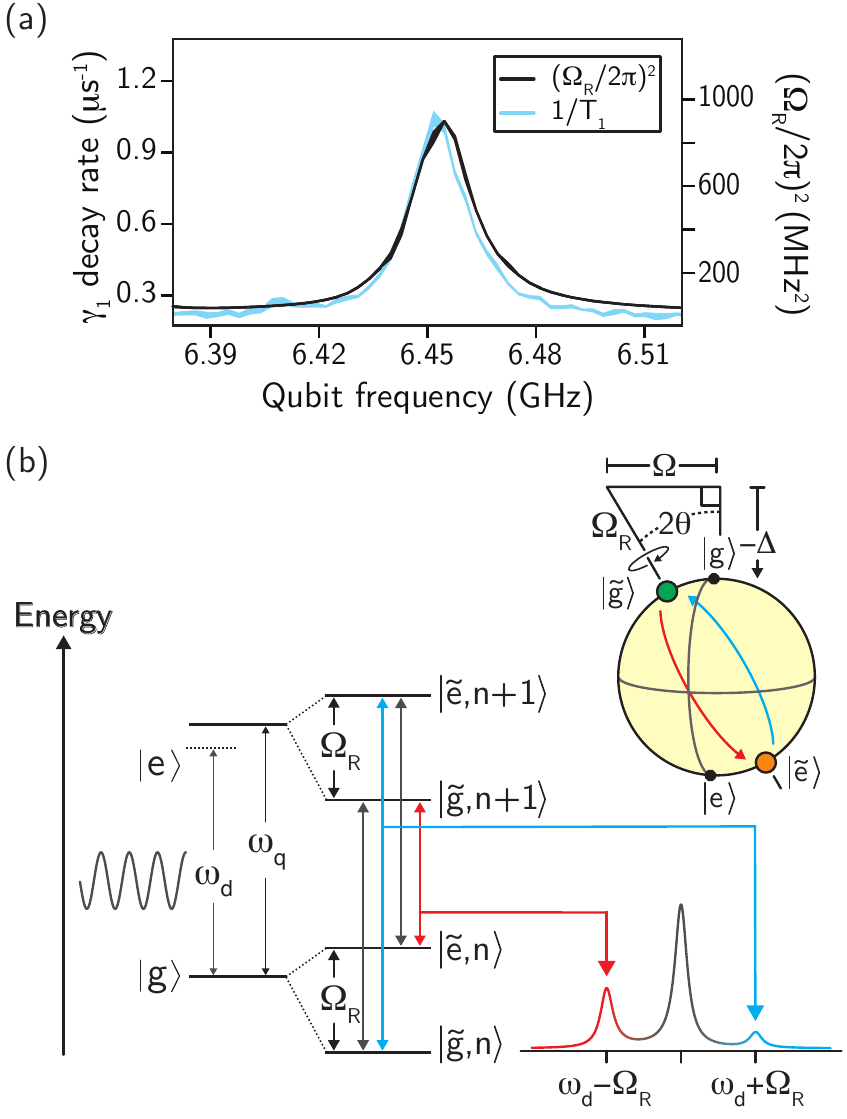}
\caption{(a) The qubit energy relaxation rate \pmh{$\gamma_1$ (\redline{cyan} trace)} and the qubit Rabi frequency squared $(\Omega_R/2\pi)^2$ (\redline{black} trace) versus the qubit resonance frequency $\omega_q/2\pi$. The qubit decay rate and Rabi frequency resulting from a resonant drive applied through the photonic crystal both indicate the frequency dependent coupling to the environment. Qubit decay rates and Rabi frequencies were extracted from state readout on time varied measurement sequences. The width of the data traces represents the standard error based on nine separate measurements. (b) The emitter, dressed by a coupling to the light-field, has an emission spectrum known as the Mollow triplet. The Mollow triplet takes on an asymmetric character in the presence of an off-resonant drive or when the local density of states of the driven emitter enhances one of the sideband transitions.}
\label{fig:fig2}
\end{figure}

Our experiment comprises a one-dimensional photonic crystal coupled to a flux tunable transmon-type superconducting artificial atom \cite{koch07,paik11} housed inside a waveguide cavity ($\omega_c/2\pi=7.801\,\text{GHz}$) (Fig.~1a) at milliKelvin temperatures. The transmon circuit has an anharmonic energy potential, allowing the two lowest energy levels to be addressed as a qubit transition. Outside the cavity, the photonic crystal is a coaxial transmission line with a spatially modulated impedance \cite{SM}, which connects the strongly coupled antenna port of the waveguide cavity to the $50\,\Omega$ electromagnetic environment of the microwave readout chain. The photonic crystal consists of 25 impedance steps ($Z_\text{LO}=30\,\Omega$, $Z_\text{HI}=50\,\Omega$) along the coaxial line, resulting in the opening of a bandgap (Fig.~1b) \cite{joan08, mirh18}. Because the transmon is dispersively coupled to the cavity, the qubit transition interacts perturbatively with the photonic crystal density of states. The decay of the transmon qubit into the photonic crystal is strongly influenced by the presence of the photonic bandgap \cite{byko72,yabl87}, since the rate of spontaneous emission is proportional to the local density of states at the transition frequency of an emitter \cite{dira27,purc46}.
 
Furthermore, the qubit-cavity dispersive coupling enables single shot readout using the Jaynes-Cummings nonlinearity technique at the bare cavity resonance \cite{reed10}. We use this to conduct full quantum state tomography of the qubit and characterize the bath engineering decay process. {Readout is performed by driving the strongly coupled port of the cavity through the photonic crystal. At a critical drive power the threshold behavior of this readout technique is observed in the phase shift of the reflection tone, achieving a readout fidelity of $\mathcal{F}=0.8$, amenable to qubit state tomography. To account for this unideal readout fidelity, we calibrate tomography measurements by preparing eigenstates of $\langle\sigma_x\rangle$, $\langle\sigma_y\rangle$, and $\langle\sigma_z\rangle$, measuring their expectation values, and rescaling experimental expectation values accordingly.}

Before implementing the bath engineering protocol, we characterize the interaction of the qubit and the photonic crystal across a range qubit transition frequencies. First, we determine the frequency dependence of the local density of states by flux biasing the qubit transition to a specific frequency and performing standard $T_1$ decay measurements (Fig.~2a). Variation of the qubit decay rate $\gamma_1=1/T_1$ is attributed to changes in the local density of states according to,
\begin{equation}\label{eq:purcell}
\gamma_1=\gamma_\mathrm{d}+\rho(\omega_q)(g/\Delta_\mathrm{q})^2\kappa,
\end{equation}
which is decay rate of the qubit dispersively coupled to a single cavity mode \cite{schu07,houc08}, where $\kappa/2\pi=18\,\text{MHz}$ is the cavity linewidth, $g/2\pi=200\,\text{MHz}$ is the qubit-cavity coupling rate, $\Delta_\mathrm{q}=\omega_c-\omega_q$ is the qubit-cavity detuning, $\rho(\omega_q)$ is the local density of states at the qubit frequency, and $\gamma_\mathrm{d}$ is the qubit decay rate into other dissipation channels. As mentioned, the cavity provides a filtered coupling to the strongly varying density of states $\rho(\omega)$ provided by the photonic crystal.

To verify the measured qubit decay is in fact influenced by the local density of states of the photonic crystal, we additionally investigate variations of the coupling rate between the qubit and its photonic crystal environment. At each flux bias, we perform resonant Rabi frequency measurements from a drive of a fixed amplitude applied through the photonic crystal (Fig.~2a). Similar to qubit decay, variation of the Rabi frequency is attributed to qubit absorption and emission rates, due to the qubit coupling to the $50\,\Omega$ continuum by way of the photonic crystal. By comparing the decay rate $\gamma_1$ and the Rabi frequency squared $(\Omega_R/2\pi)^2$, we find agreement in the proportional changes of decay and coupling rates and establish that the photonic crystal forms the spectral density of states for qubit emission and absorption, {since both the qubit decay rate $\gamma_1$ and the squared Rabi frequency both depend proportionally on the local density of states.} From this, we attribute changes of the qubit decay rate to the large variation of the local density of states between the stop-band and pass-band of the photonic crystal.

We now apply a coherent drive on the qubit {through the weakly coupled cavity port} to implement our bath engineering protocol. The coherent drive, along with the photonic crystal spectral density of states, determines the steady-state of the bath engineering process by inducing specific decay transitions of the qubit {\cite{yan13}}. We solve for this steady-state by considering the system dynamics under drive and decay. The time-evolution of our bath engineering process is simply modeled as a two level emitter under coherent drive, since the qubit interacts only weakly with the dissipative photonic states \cite{SM}. The open quantum system dynamics of a transmon qubit driven from a coherent tone at frequency $\omega_d$ at resonant Rabi frequency $\Omega$ and detuning $\Delta=\omega_d-\omega_q$, can be described by the Lindblad master equation \cite{lind76}, assuming timescales of photonic state relaxation are much shorter than that of the qubit  \cite{scal07,gamb08,bois09,aron16,liu17},
\begin{align}\label{eq:lind}
\dot{\rho}&=
i[\rho,H]+\gamma_0\cos(\theta)\sin(\theta)\mathcal{D}[\tilde{\sigma}_z]\rho+\nonumber\\
&\hspace{1cm}\gamma_-\sin^4(\theta)\mathcal{D}[\tilde{\sigma}_+]\rho+\gamma_+\cos^4(\theta)\mathcal{D}[\tilde{\sigma}_-]\rho
\end{align}
where $\rho$ is the reduced density operator for the qubit dressed by the light-field in the rotating frame of the drive. The operators $\tilde{\sigma}_\pm$ are Pauli raising and lowering operators in the qubit dressed basis and $H=\Omega_R\tsz/2$ is the Hamiltonian of the qubit with $\hbar=1$, and the generalized Rabi frequency $\Omega_R=\sqrt{\Omega^2+\Delta^2}$. The dephasing rate in the dressed basis is $\gamma_0$ and transitions $\tilde{\sigma}_\pm$ between these eigenstates occur at rates $\gamma_\mp$, as captured by the dissipation superoperator $\mathcal{D}[L]\rho=2(L\rho L^\dagger-L^\dagger L\rho-\rho L^\dagger L)/2$. The dressed qubit energy eigenbasis $\{\ket{\tilde{g}},\ket{\tilde{e}}\}$ formed by coherent driving at frequency $\omega_d$, comprises superpositions of the bare qubit eigenstates $\{\ket{g},\ket{e}\}$ as illustrated in Figure~2b,
\begin{align*}
\ket{\tilde{g}} &=\cos(\theta)\ket{g} - \sin(\theta)\ket{e}\\
\ket{\tilde{e}} &=\sin(\theta)\ket{g} + \cos(\theta)\ket{e},
\end{align*}
where $\tan(2\theta)=-\Omega/\Delta$.  After sufficiently long time evolution of Eq.~\ref{eq:lind}, the qubit relaxes to a nonequilibrium effective ground state: a steady-state of the driven-dissipative dynamics \cite{raft14}. Importantly, a superposition state results from an asymmetry in transition rates $\tilde{\sigma}_\pm$ between the dressed states, due to the frequency dependence of the photonic density of states.

We now discuss the role of the photonic crystal for decay to a specific qubit steady-state. Without drive, the qubit decays to its ground state according to Eq.~(\ref{eq:lind}) which reduces to $\dot{\rho}=\gamma_1\mathcal{D}[\sm]\rho$, where $\gamma_1$ is given by Eq.~(\ref{eq:purcell}) and the qubit emits a photon at the qubit transition frequency $\omega_q$. A coherent drive on the qubit introduces a new energy scale $\Omega$, allowing the qubit to couple to the electromagnetic field at frequencies $\omega_d$ and \pmh{$\omega_d \pm \Omega_R$}. For a resonant drive, the latter frequencies correspond to transitions given by the operators, 
$$
\begin{array}{cc}
\tsm = \ket{+x}\bra{-x}, & \tsp=\ket{-x}\bra{+x},
\end{array}
$$
in the limit of strong excitation where $\Omega\gg\gamma_\pm$. If the spectral density of states of the electromagnetic field at $\omega_d \pm \Omega$ are equal, the rates for the transitions $\tilde{\sigma}_\pm$ are equally favored resulting in a  mixed state for the qubit. However, when the photonic crystal results in a colored density of states, the rates $\gamma_\mp$ associated with $\tilde{\sigma}_\pm$ can be unequal, favoring decay to either the $\ket{+x}$ or $\ket{-x}$ state.

Emission of the driven system creates field correlations that manifest as the Mollow triplet spectrum (Fig.~2b) \cite{baur09,toyl16,moll69}. The asymetry in the rates $\gamma_\mp$ leads to an asymmetry in the Mollow triplet. In the ideal scenario for bath engineering, there is a thoroughly dissimilar local density of states at frequencies \pmh{$\omega_d \pm \Omega_R$}, resulting in a single sideband Mollow triplet and deterministic decay to one of the two dressed states.

\begin{figure}
\centering
\includegraphics[width=8.6cm]{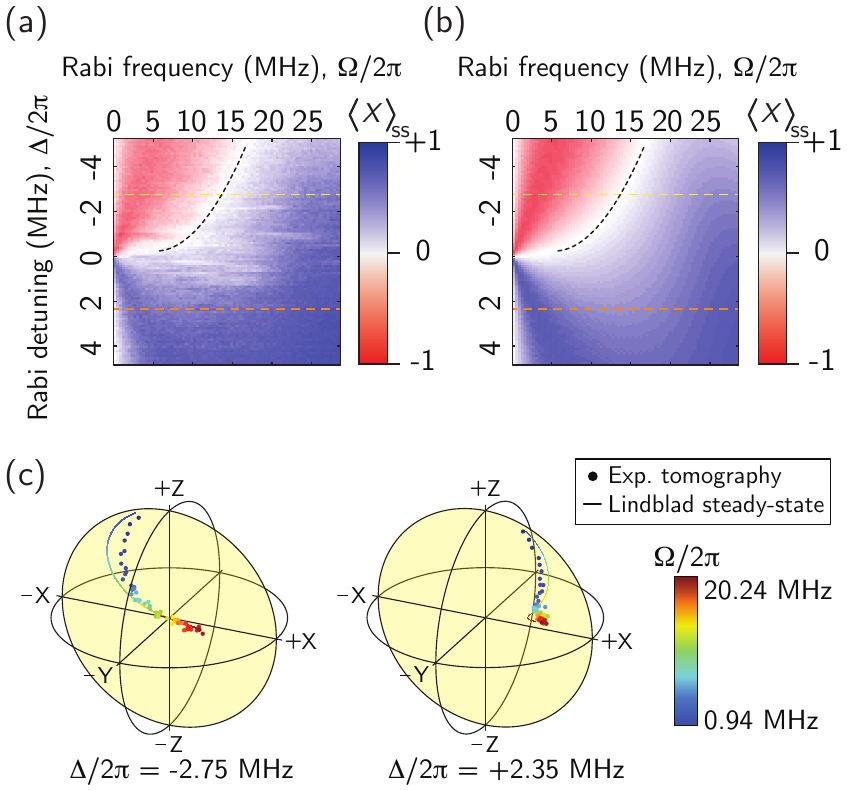}
\caption{(a) The steady-state coherence \pmh{$\langle X\rangle_\mathrm{ss}$ for a range of Rabi drive amplitudes and detunings. The black dashed line indicates drive parameters that give a steady-state of zero coherence determined by the decay rates given by measurements in Figure 2a {and Eq.~\ref{eq:db}}}. (b) The calculated $\langle X\rangle_\mathrm{ss}$ based on Eq.~\ref{eq:lind} and the experimental driving parameters. The black dashed line indicates the same drive parameters as in panel (a). (c) The Bloch sphere representation of the calculated and measured qubit steady-state, $\vec{r}=\mathrm{Tr}(\rho_\mathrm{ss} \vec{\sigma})$, for {the drive detuning $\Delta/2\pi=-2.75\,\text{MHz}$ (yellow dashed line in panels (a) and (b)), and $\Delta/2\pi=2.35\,\text{MHz}$ (orange dashed line in panels (a) and (b))}.}
\label{fig:3a}
\end{figure}

We demonstrate bath engineering decay to a dressed state by flux tuning the qubit to $\omega_q/2\pi=6.4766\,\text{GHz}$ where the local density of states varies dramatically, as shown in Figure~2a. In Figure~3a, we display the measured steady-state qubit coherence $\langle X\rangle_\mathrm{ss} \equiv \mathrm{Tr}(\rho_\mathrm{ss} \sigma_x)$, where $\rho_\mathrm{ss}$ is the tomographically reconstructed qubit state after $15.95\,\mu\text{s} \gg 1/\gamma_\pm$ of driving and $\sigma_x$ is the Pauli operator in the undressed basis. Here, we observe two signatures of the photonic crystal density of states. First, we find the steady-states mapped in Figure~3a,b contain a feature of zero coherence (dashed-line) for certain coherent drive parameters of detuning and amplitude. This occurs when the two terms $\gamma_\pm$ in Eq.~(2) cancel due to the dependence of both $\theta$ and $\gamma_\pm$ on $\Delta$ and $\Omega$. A maximally mixed steady-state is a consequence of equal transition rates between dressed states. Physically, the overlap of the dressed states with the globally favored ground state competes with the dressed state favored by $\gamma_\pm$. In a picture of detailed balance for the rate of transitions between dressed states, this occurs for drive parameters satisfying the relation
\begin{equation}\label{eq:db}
\gamma_-(\Omega,\Delta)\sin^4(\theta)=\gamma_+(\Omega,\Delta)\cos^4(\theta),
\end{equation}
which was used to calculate the dashed lines of Figure 3a,b. A second signature of the photonic crystal is observed by the increase of the steady-state coherence for a resonant drive. Although this coherence is limited in our experiment by decay to other dissipation channels, we find an overall increase of steady-state coherence because the dressed state transition rates become more asymmetric as the Mollow triplet spectrum widens in the presence of a colored local density of states. While small coherences can be created from a weak drive in resonance fluorescence \cite{carm87}, the observation of coherence from a strong drive is a clear indicator of an asymmetry in the rates $\gamma_\pm$ 
{due to the density of state of the photonic crystal. Furthermore, we note that the asymmetric density of states of the readout cavity is negligible due to its large detuning from the qubit resonance $\Omega,\Delta\ll\Delta_\mathrm{q}$.}

Consequently, we find that the qubit is ``cooled'' to a chosen superposition state in the eigenbasis of the undriven qubit from a proper selection of a drive phase, frequency, and amplitude (Fig.~3), enabled by the asymmetric density of states of the photonic crystal. The theory colormap of Figure~3b was produced by solving for the steady-states of Eq.~\ref{eq:lind} given the local density of states as inferred from measurements shown in Figure~2. This theory reproduces all qualitative features of the tomography results and has quantitative agreement when including additional pure dephasing of the qubit transition $\gamma_\phi=0.029\,\mu\text{s}^{-1}$, consistent with typical limits of coherence for transmon qubits \cite{SM}.

In conclusion, we have shown that the driven and dissipative dynamics of a transmon qubit weakly coupled to a photonic crystal can be used for quantum bath engineering, as we have verified with full state tomography. Our protocol robustly prepares a desired qubit superposition state, realized as an effective ground state of the driven-dissipative system. The colored density of states introduced from the photonic crystal is crucial for our method and highlights impedance engineering of the electromagnetic environment as a key aspect of bath engineering for circuit quantum electrodynamics. In future bath engineering implementations, the photonic density of states can be tailored by fabrication techniques with lumped element metamaterials \cite{mirh18} and \textit{in situ} tunability of coupling rates between photonic modes \cite{lu17,coll18}. Additionally, quantum monitoring of dissipative photonic modes of the environment can further the scope of bath engineering protocols for non-unitary heralding of quantum states and quantum control by dynamical feedback \cite{sayr11,vija12,shan13,rist13,lang14,roch14,nagh16,camp16}.

\begin{acknowledgements}
{We thank A.~A. Clerk, P. Bertet, I. Martin, and J. Monroe for helpful conversations, and we appreciate J.~R. Lane and J. Pollanen for preliminary experimental contributions.} We acknowledge research support the NSF (Grant PHY-1607156) and from ONR (Grant 12114811). This research used facilities at the Institute of Materials Science and Engineering at Washington University. D.~T. acknowledges support from the Rigetti Computing Postdoctoral Fellowship.
\end{acknowledgements}

\section*{Appendix A: Photonic crystal fabrication and characterization}
The photonic crystal was hand fabricated from a $50\,\Omega$ semi-rigid transmission line (Micro Coax UT-085C-TP-LL). Since the TEM propogation mode geometry determines the characteristic impedance of the transmission line, a periodic modulation of the transmission line geometry along the line forms a finite length one-dimensional photonic crystal. Transmission line sections were mechanicaly deformed by crushing the coax, creating lengths of characteristic impedance $Z_0\simeq30\,\Omega$, as found consistent with Ansys HFSS simulation.

{We modeled the photonic crystal in AWR Microwave Office as a Chebychev Type I bandstop filter. Given prior knowledge that squashed SMA sections have $\simeq30\,\Omega$ characteristic impedance and the dielectric constant of PTFE ($\epsilon\simeq2$), we optimized for an experimentally convenient frequency for the upper band edge, resulting the the parameters given in Table~I. The lengths of the 25 impedance sections were used to fabrication the photonic crystal. In Figure~4, we present measured and calculated scattering parameters for the photonic crystal. The scattering parameters were calculated from cascaded $ABCD$ transfer matrices of transmission line sections of length in Table~I and with a minor adjustment to the transmission line dielectric constant ($\epsilon=1.96$).}
\begin{widetext}
\begin{center}
\begin{table}[!htbp]\label{table:lengths}
\begin{tabular}{|c|c	|	c	||	c	|	c	|	c	||	c	|	c	|	c	||	c	|	c	|	c	||	c	|	c	|	c	|}\hline
Step \#	&	$Z_0\,(\Omega)$	&	$\ell$ (mm)	&	Step \#	&	$Z_0\,(\Omega)$	&	$\ell$ (mm)	&	Step \#	&	$Z_0\,(\Omega)$	&	$\ell$ (mm)	&	Step \#	&	$Z_0\,(\Omega)$	&	$\ell$ (mm)	&	Step \#	&	$Z_0\,(\Omega)$	&	$\ell$ (mm)	\\	\hline
1	&	30	&	9.1	&	6	&	50	&	9.4	&	11	&	30	&	10.2	&	16	&	50	&	9.7	&	21	&	30	&	9.7	\\	\hline
2	&	50	&	9.4	&	7	&	30	&	9.9	&	12	&	50	&	9.7	&	17	&	30	&	10.2	&	22	&	50	&	10.9	\\	\hline
3	&	30	&	9.1	&	8	&	50	&	9.7	&	13	&	30	&	10.2	&	18	&	50	&	9.7	&	23	&	30	&	9.1	\\	\hline
4	&	50	&	10.7	&	9	&	30	&	10.2	&	14	&	50	&	9.7	&	19	&	30	&	9.9	&	24	&	50	&	9.4	\\	\hline
5	&	30	&	9.7	&	10	&	50	&	9.7	&	15	&	30	&	10.2	&	20	&	50	&	9.4	&	25	&	30	&	9.1	\\	\hline
\end{tabular}
\caption{The phontonic crystal filter was fabricated by creating a modulation of the characteristic impedance $Z_0$ of a transmission line. The photonic crystal is modelled as a Chebychev Type I bandstop filter. The lengths of the impedance sections were informed from filter simulations in AWR Microwave Office.}
\end{table}
\end{center}
\end{widetext}

\begin{figure}[H]\label{fig:PCmodel}
\centering
\includegraphics[width=8.6cm]{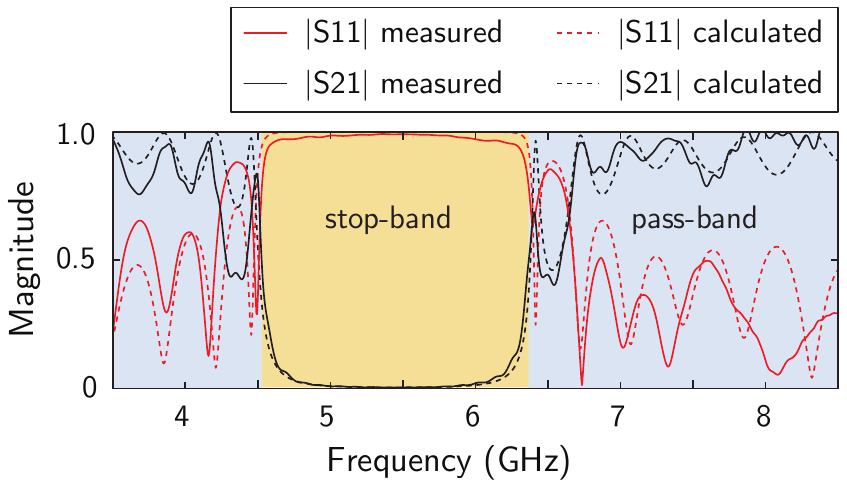}
\caption{The scattering parameters of the photonic crystal filter were measured from a two-port $50\,\Omega$ calibrated vector network analyzer. We compare the measured reflection (S11) and transmission (S21) scattering parameters to those calculated from cascaded $ABCD$ transfer matrices of transmission line sections with parameters given in Table~I.}
\end{figure}

\section*{Appendix B: Additional data of the photonic crystal characterization}
The photonic crystal density of states was characterized from both unitary and dissipative dynamics of the qubit. By the same methods as described for Figure~2 of the main text, we show the qubit decay rate is determined by the environment of the photonic crystal density of states.
\begin{figure}[h]
\centering
\includegraphics[width=8.6cm]{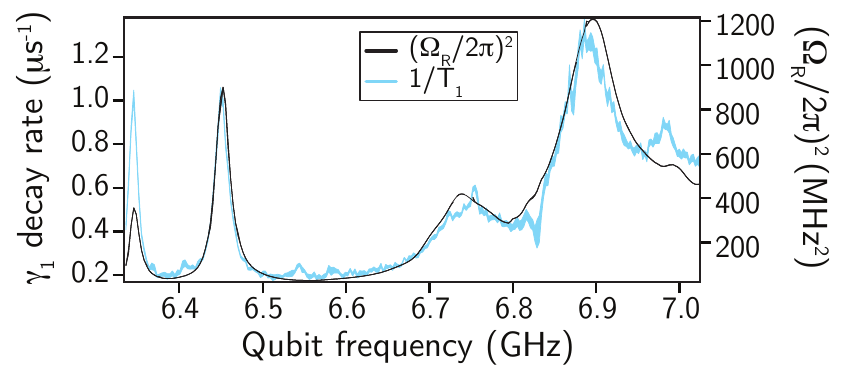}
\caption{{The photonic crystal density of states was characterized across a large frequency band by flux tuning the qubit and measuring the qubit lifetime (\redline{cyan} trace) and squared Rabi frequency (\redline{black} trace) at each qubit transition frequency.}}
\label{fig:suppfig4}
\end{figure}

\section*{Appendix C: Steady-state simulation}
A qubit with an energy eigenbasis $\{\ket{g},\ket{e}\}$ is described by the Hamiltonian $H=-\omega_q\sigma_z/2$ in the lab frame. When the qubit is dipole coupled to a coherent drive of frequency \redline{$\omega_d$, we} transform the lab frame Hamiltonian of the driven qubit $H=-\omega_q\sz/2+\Omega\sx\cos(\omega_d t)$ into the rotating frame of the drive with the unitary operator $U=e^{-i\omega_d t\sz/2}$ as $H\rightarrow UHU^\dagger+i\dot{U}U^\dagger$. The rotating frame Hamiltonian is
\begin{equation}\label{eq:h1}
H_q=\frac{\Delta}{2}\sigma_z+\frac{\Omega}{2}\sigma_x
\end{equation}
upon neglecting rapidly oscillating terms and where $\Delta=\omega_d-\omega_q$ is the qubit-drive detuning and $\Omega$ is the frequency of Rabi oscillations in the case of a resonant drive. We diagonalize Eq.~\ref{eq:h1} to find the dressed energy eigenstates,
\begin{align*}
\ket{\tilde{g}} &=\cos(\theta)\ket{g} - \sin(\theta)\ket{e}\\
\ket{\tilde{e}} &=\sin(\theta)\ket{g} + \cos(\theta)\ket{e},
\end{align*}
where $\tan\,2\theta=-\Omega/\Delta$ and $0\le\theta<\pi/2$. We define the energy eigenstates such that $\ket{\tilde{g}}\simeq\ket{g}$ when the qubit is driven far-red detuned and $\ket{\tilde{e}}\simeq\ket{g}$ when the qubit is driven far-blue detuned. We rewrite the Hamiltonian in the dressed state basis as,
$$
H_q=\frac{\Omega_R}{2}\tsz
$$
where $\Omega_R=\sqrt{\Omega^2+\Delta^2}$ and $\tilde{\sigma}_z=\sin(2\theta)\sigma_x-\cos(2\theta)\sigma_z$.

We wish to consider the interaction picture of the driven qubit weakly coupled to a dissipative environment, such that we can treat the interaction as a perturbation. The driven qubit interacting with dissipative modes of the electromagnetic environment is described by the interaction Hamiltonian in the rotating frame of the drive,
$$
H_{int}=\sum_k g_k (\sm\bdk e^{i\Delta_k t}+h.c.)
$$
where $g_k$ is the coupling strength to the electromagnetic mode of frequency $\omega_k=\Delta_k-\omega_d$ with creation operator $\bdk$. Each term of the interaction Hamiltonian can be expressed in terms of dressed state operators as,
\begin{align*}
H_{int}^k&=g_k(\cos^2(\theta)\tsm-\sin^2(\theta)\tsp\\
&\hspace{1cm}\ldots+\sin(\theta)\cos(\theta)\tsz)\bdk e^{i\Delta_k t}+h.c.
\end{align*}
where we have simply made the substitution $\sm=\cos^2(\theta)\tsm-\sin^2(\theta)\tsp+\sin(\theta)\cos(\theta)\tsz$. We transform both the qubit and interaction Hamiltonian into the rotating frame of the dressed qubit described by the transformation $H\rightarrow UHU^\dagger+i\dot{U}U^\dagger$ where $U=\exp(i\Omega_R\tsz/2)$, giving the Hamiltonian
\begin{align*}
H(t)&=\sum_k g_k(\cos^2(\theta)\tsm e^{i(\Delta_k+\Omega_R)t}\\
&\hspace{1cm}\ldots-\sin^2(\theta)\tsp e^{i(\Delta_k-\Omega_R)t}\\
&\hspace{1cm}\ldots+\sin(\theta)\cos(\theta)\tsz  e^{i\Delta_kt})\bdk+h.c.
\end{align*}
As we consider the time evolution of both the qubit and dissipative environment in the interaction picture, we assume the environment modes are sufficiently dissipative, such that we can make the Born approximation and trace out the environment degrees of freedom. We subsequently make the Markov approximation, and assume time evolution is coarse grained enough for the environment local density of states to determine jump rates of the open system dynamics. The time evolution for the reduced density matrix of the qubit is described by the Lindblad master equation,
\begin{align*}
\dot{\rho}&=\gamma_-\mathcal{D}[\sin^2(\theta)\tsp]\rho+\gamma_+\mathcal{D}[\cos^2(\theta)\tsm]\rho\\
&\hspace{1cm}\ldots+\gamma_0\mathcal{D}[\sin(\theta)\cos(\theta)\tsz]\rho.
\end{align*}
where $\mathcal{D}[L]\rho=2(L\rho L^\dagger-L^\dagger L\rho-\rho L^\dagger L)/2$, $\gamma_-=2\pi\sum_k g_k^2\delta(\omega_k-(\Delta_k-\Omega_R))$, $\gamma_+=2\pi\sum_k g_k^2\delta(\omega_k-(\Delta_k+\Omega_R))$, and $\gamma_0=2\pi\sum_k g_k^2\delta(\omega_k-\Delta_k)$. 

Numerical calculations were performed in the dressed state basis including unitary evolution from Rabi oscillations described by the master equation,
\begin{align*}
\dot{\rho}&=i[\rho,\Omega_R\tsz/2]+\gamma_-\mathcal{D}[\sin^2(\theta)\tsp]\rho+\gamma_+\mathcal{D}[\cos^2(\theta)\tsm]\rho\\
&\hspace{1cm}\ldots+\gamma_0\mathcal{D}[\sin(\theta)\cos(\theta)\tsz]\rho+\frac{\gamma_\phi}{2}\mathcal{D}[\sz]\rho.
\end{align*}
where the final term of the master equation captures an additional pure dephasing of rate $\gamma_\phi$ in the lab frame of the qubit.
The density matrix time evolution was numerically solved by recasting the Lindblad superoperator into a $4\times4$ matrix which maps a vector representation of the density matrix to another vector. The qubit density matrix is expressed as the column vector, $\vec{\rho}=(\rho_{gg},\,\rho_{ge},\,\rho_{eg},\, \rho_{ee})^T$. We construct the Lindblad operator $\mathcal{L}$ as a matrix in operator space by expressing left-operation ($A\rho$) and right-operation ($\rho A$) on the density matrix with tensor products. Matrices of left- and right- operation are
$$
A\rho\rightarrow (\mathbb{I}\otimes A)\vec{\rho}=
\begin{pmatrix}
A_{11} & A_{12} & 0 & 0\\
A_{21} & A_{22} & 0 & 0\\
0 & 0 & A_{11} & A_{12}\\
0 & 0 & A_{21} & A_{22}\\
\end{pmatrix},
$$
and,
$$
\rho A\rightarrow (A\otimes\mathbb{I})\vec{\rho}=
\begin{pmatrix}
A_{11} & 0 & A_{12} & 0\\
0 & A_{11} & 0 & A_{12}\\
A_{21} & 0 & A_{22} & 0\\
0 & A_{21} & 0 & A_{22}\\
\end{pmatrix}.
$$
Time evolution from an initial qubit state is calculated from the equation, $\vec{\rho}(t) = e^{\mathcal{L}t}\vec{\rho}(0)$, where we perform matrix exponentiation of $\mathcal{L}t$ by finding the matrix $V$ which diagonalizes the Lindblad matrix. After converting the density matrix vector into a matrix operator ($\vec{\rho}(t)\rightarrow\rho_t$), we then calculate expectation values in the lab frame rotating with the drive,
$$
\begin{array}{ccc}
\langle X\rangle_t=\mathrm{tr}(\sx\rho),\,&
\langle Y\rangle_t=\mathrm{tr}(\sy\rho),\,&
\langle Z\rangle_t=\mathrm{tr}(\sz\rho),
\end{array}
$$
where \redline{the Pauli operators in terms of the dressed state basis are} $\sx=\cos(2\theta)\tsx-\sin(2\theta)\tsz$, $\sy=\tsy$, and $\sz=\sin(2\theta)\tsx+\cos(2\theta)\tsz$.

\section*{Appendix D: Full state tomography}
\begin{figure}[H]
\centering
\includegraphics[width=8.6cm]{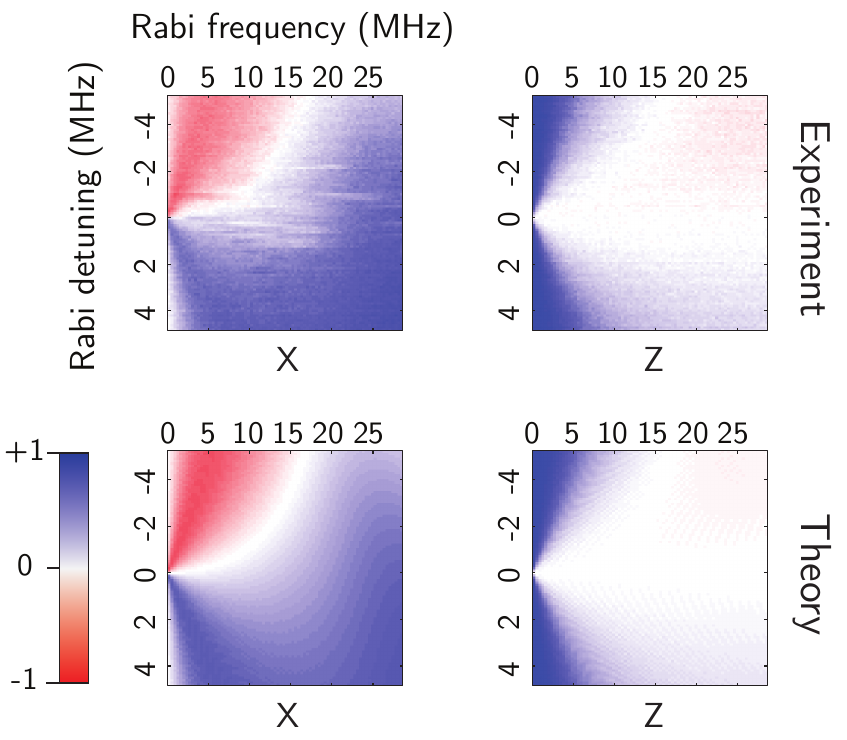}
\caption{Tomographic measurements of the qubit state after $15.95\,\mu\text{s}$ of drive, as presented in Figures~3 of the main text. The phase of tomography pulses were chosen to be in the rotating frame of the drive, thus $\langle Y\rangle_\mathrm{ss}=0$ and all states lie in the $X$-$Z$ plane of the Bloch sphere.}
\end{figure}

\section*{Appendix E: Experiment setup}
\begin{figure}[H]
\centering
\includegraphics[width=8.6cm]{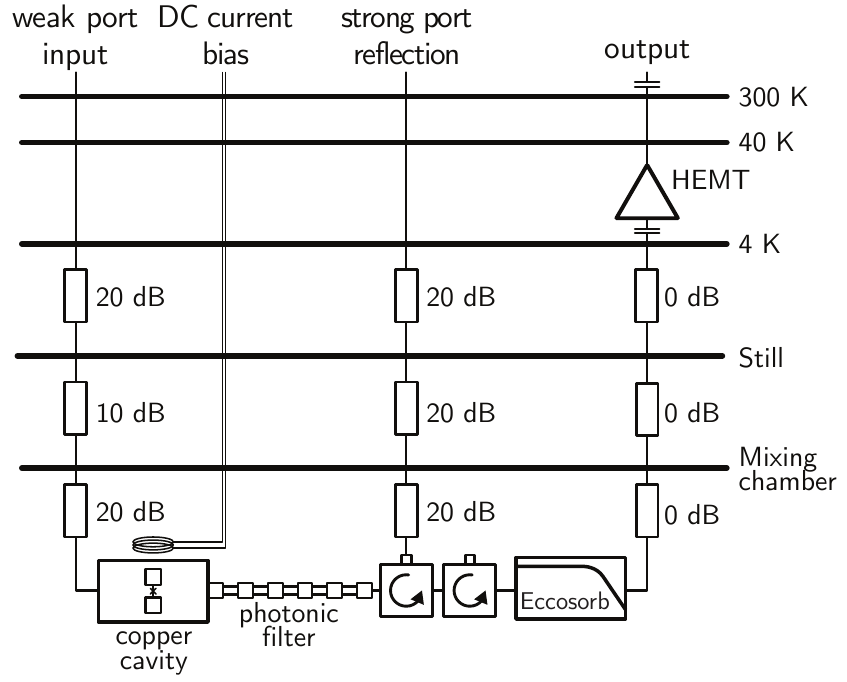}
\caption{The cryogenic microwave schematic of the experiment.}
\end{figure}
\end{document}